# Natural Computational Architectures for Cognitive Info-Communication

**Gordana Dodig-Crnkovic**[1, 2]

[1] Department of Computer Science and Engineering, Chalmers University of Technology, Gothenburg, Sweden

[2] Division of Computer Science and Software Engineering, School of Innovation, Design and Engineering, Mälardalen University, Västerås, Sweden

dodig@chalmers.se

## Abstract

Recent comprehensive overview of 40 years of research in cognitive architectures, (Kotseruba and Tsotsos 2020), evaluates modelling of the core cognitive abilities in humans, but only marginally addresses biologically plausible approaches based on natural computation. This mini review presents a set of perspectives and approaches which have shaped the development of biologically inspired computational models in the recent past that can lead to the development of biologically more realistic cognitive architectures. For describing continuum of natural cognitive architectures, from basal cellular to human-level cognition, we use evolutionary info-computational framework, where natural/ physical/ morphological computation leads to evolution of increasingly complex cognitive systems. Forty years ago, when the first cognitive architectures have been proposed, understanding of cognition, embodiment and evolution was different. So was the state of the art of information physics, bioinformatics, information chemistry, computational neuroscience, complexity theory, self-organization, theory of evolution, information and computation. Novel developments support a constructive interdisciplinary framework for cognitive architectures in the context of computing nature, where interactions between constituents at different levels of organization lead to complexification of agency and increased cognitive capacities. We identify several important research questions for further investigation that can increase understanding of cognition in nature and inspire new developments of cognitive technologies. Recently, basal cell cognition attracted a lot of interest for its possible applications in medicine, new computing technologies, as well as micro- and nanorobotics. Bio-cognition of cells connected into tissues/organs, and organisms with the group (social) levels of information processing provides insights into cognition mechanisms that can support the development of new AI platforms and cognitive robotics.

## 1 INTRODUCTION

In 1958 John von Neumann wrote "The computer and the brain" (von Neumann 1958)- the book describing information processing architecture of computers as based on then-current understanding of brain organization, with separate memory, input/output unit, arithmetic/logic unit, and a control unit. Von Neumann architecture is still in use. However, understanding of the brain have changed radically, and we may hope that new understanding of the brain and cognition will bring about new computational architectures.

In an overview of 40 years of research and practical applications in cognitive architectures, (Kotseruba and Tsotsos 2020) address the adequacy of cognitive architectures in modelling of the core cognitive abilities in humans, including perception, attention, action, memory, learning, and reasoning. Apart from presenting the state-of-the-art of the research through 84 human-level cognitive architectures, authors briefly mention deep learning, and why it does not qualify as *unified model of cognition in humans*. We will come back to recent developments in deep learning.

This mini-review gives a framework not fully presented in (Kotseruba and Tsotsos 2020), of natural computational cognitive architectures for info-communication. In this naturalistic approach, the underlying assumption is that cognition in nature is a manifestation of biological processes (that subsume chemical and physical levels) in living organisms (Maturana and Varela 1992; Stewart 1996; Dodig-Crnkovic 2007; Jagers op Akkerhuis 2010; Lyon 2005; 2015; Lyon and Kuchling 2021) from single cells to humans. Recent work of (Piccinini 2020) addresses biological cognition in organisms with nervous systems as result of neurocomputation. However, "*cognitive operations we usually ascribe to brains—sensing, information processing, memory, valence, decision making, learning, anticipation, problem solving, generalization and goal directedness—are all observed in living forms that don't have brains or even neurons.*" (Levin et al. 2021)

Based on empirical and theoretical insights about cognition and its evolution and development in nature (Walker, Davies, and Ellis 2017) (Dodig-Crnkovic 2017), from basal/ basic/ primitive/ elementary/ cellular to complex form of human cognition, (Manicka and Levin 2019; Levin et al. 2021; Lyon et al. 2021; Dodig-Crnkovic 2020; Stewart 1996; Dodig-Crnkovic 2014a) modelled on natural information processing (natural computation), we identify several cognitive architecture topics that deserve more study.

The rest of the review is organized as follows. Section 2 presents info-computational cognitive architectures in natural (biological) systems, while Section 3 is focusing on the natural info-computation processes as a basis of natural cognitive architectures. Section 4 addresses naturalizing communication, while Section 5 highlights centrality of time aspect in cognitive architecture. Section 6 provides a discussion and several open questions of natural cognitive architectures worth further study. Section 7 offers a conclusion.

## 2  NATURAL INFO-COMPUTATIONAL COGNITIVE ARCHITECTURES

An info-computational natural cognitive architecture is a natural informational structure that generates cognition, in which knowledge and skills of a cognitive agent are embodied. Through natural computational processes it produces intelligent behavior of natural systems in complex environments. Traditionally cognition has been studied as a result of brain activity in humans. There was an opposition between understanding of brain function as a result of symbol processing of Turing computation type (Wells 2004), distributed computation models (Clark 1989), and dynamic models of a brain as a meta-stable oscillating system (Kelso, Dumas, and Tognoli 2013). It is important to realize that all three "modes operandi" in the brain can suitably be modelled as natural computations (natural information processing) on different levels of organization.

Let us start by introducing the framework for naturalizing cognition, with two basic elements: natural (embodied) information, and natural computation.

# Natural Computational Architectures

## Naturalized Cognition. Thinking Fast and Slow - System 1 and System 2

Cognitive architectures started as a research field with the goal to model *human mind* and build *human-level artificial intelligence*. By connecting models and mechanisms with observed cognitive/intelligent behaviors, they contribute to cognitive science and AI. However, cognition in nature appears throughout biological systems (Baluška and Levin 2016; Lyon 2005; 2015; Almér, Dodig-Crnkovic, and von Haugwitz 2015; Lyon et al. 2021) and it is important to understand its evolutionary development from the basal/basic/elementary cognition to the human level (Manicka and Levin 2019; Levin et al. 2021). Thus, cognitive architectures have considerable explanatory and practical value.

In the context of the (Kotseruba and Tsotsos 2020) systematization, info-computational naturalist cognitive architecture is a hybrid (connectionist-symbolic), biologically realistic conceptual framework aiming to integrate current knowledge from variety of research fields, such as cognitive science, computational neuroscience, bioinformatics, computability, biology, and new evolutionary synthesis.

This approach is based on the view of hierarchical recursive structure of information processing in nature, which is especially important for living organisms, from cells, to tissues, organs, organisms and their groups – all of them communicating at different levels of organization by exchanging specific types of information – physical (elementary particles, electro-magnetic), chemical (electric, molecular), biological, and symbolic.

From the time when first cognitive architectures have been proposed until now, a lot has changed in our understanding of cognition, embodiment, functioning of the brain, neurons and neuronal networks. Within AI, the field of artificial neural networks with deep learning have made an impressive progress in modeling perception on the level of data/signal processing.

However, *in humans*, two basic cognitive systems have been recognized, System 1 (reflexive, non-conscious, automatic, intuitive information processing, which is fast) and System 2 (reflective, conscious, reasoning and decision making, which is slow) (Kahneman 2011; Tjøstheim et al. 2020). As Kahneman explains, System 1 and System 2 stand for informational processes that are functional abstractions, not the brain regions.

As deep learning models are inspired by the human brain information processing, recent advances in understanding of natural cognitive systems can contribute both to better explanatory models and to future developments of constructive engineered cognitive architectures.

Deep learning level corresponds to Kahneman's fast, intuitive System 1 (Kahneman 2011), and current developments in AI are continuing towards even more ambitious goals of modelling System 2 symbolic reasoning (Russin, O'Reilly, and Bengio 2020).

It has long been recognized that mechanisms of cognition based on natural computation are far more sophisticated than the machine-like classical computationalist models based on abstract symbol manipulation (Kampis 1991). They conform to the view expressed by (Witzany 2000) and (Witzany and Baluška 2012b) that *rule-based machines are not good enough models of natural cognition which appears in highly complex living organisms*.

*Embodiment is the fundamental feature of cognition*, which implies that valence, affect, feelings and emotions must be taken into account as constitutive elements in the models of cognition (Damasio 1999; Watanabe, Hofman, and Toru 2017; Dodig-Crnkovic 2017; Dodig-Crnkovic and Giovagnoli 2017; Lyon and Kuchling 2021) and they affect both System 1 and System 2 information processing.

## 2.1   Information Processing in Embodied and Extended Cognition

Naturalized cognition as a systemic perspective means broadening the scope *beyond neurocentrism* (Cowley and Vallée-Tourangeau 2017; Gabriel 2017; Lyon 2005; Lyon et al. 2021; Levin et al. 2021) to networks of networks of information-processing agents, down to molecular level as computational basis of distributed cognition. The physical mechanism of natural computation is morphological computation (Pfeifer, Iida, and Gomez 2006; Dodig-Crnkovic 2013a; 2017), where morphology refers to form, shape, structure which defines interactions (Dodig-Crnkovic 2014b).

Cognitive architectures often identify cognition with information processing in the brain. However, with the rise of embodied cognition, neurocentrism is being challenged by a systemic view of cognition, where body substantially shapes cognitive functions. As (Cowley and Vallée-Tourangeau 2017) show, living beings also connect their bodies with artifacts, in a sense of extended cognition (Clark 2008). The crucial importance of interaction we learn as well from (Ginsburg and Jablonka 2019) view of the evolution of "the sensitive soul", providing the naturalized communication between an agent and its environment, that changes both.

## 2.2   Evolutionary View of Cognition in Nature. Scaling from Basal to Complex

If we want to learn how cognition functions in human as the most complex living organism, it is instructive to see how this ability developed through evolution, resulting in variety of cognitive architectures of organisms from bacteria to humans (Ginsburg and Jablonka 2019)(Manicka and Levin 2019) (Lyon et al. 2021).

In a naturalist approach, cognition in any living organism is a result of embodied processes that make the organism alive (Maturana and Varela 1992), or as (Stewart 1996) puts it, "Cognition = Life". Here life includes capability of growth and reproduction. All living systems are cell-based, from unicellular to complex ones, with cells organized in tissues and organs, where each cell possesses cognition. Groups of organisms like swarms and flocks exhibit social cognition. Cognitive capacities on different scales make living system goal-directed, robust and adaptive. In biomimetic (nature-inspired) robotic systems, cognition is represented by the equivalents of living functions, implemented in a robot provided with sensors, actuators and information processing units. This basal level of cognition can be of interest, as robots do not always need human-level abilities to perform their tasks (Brooks 1991). Cognition of a different, non-human type can be adequate in biomimetic soft robots (Joyee et al. 2020). Levin et al. (Levin 2020; 2019; Baluška and Levin 2016; Manicka and Levin 2019; Levin et al. 2021) present a variety of mechanisms of basal cognition where robust adaptive information processing and behavior can be used to develop new computational techniques in biological and engineered systems.

In the naturalized, evolutionary concept of cognition, development goes from the simplest organizational form of a single cell, as "cellular mind", up to the brain as "the society of mind" (Minsky 1986). In this process, body plays a vital role in shaping minds (Pfeifer and Bongard 2006). Organisms learn about the world by means of information exchanges/communication (Terzis and Arp 2011; Dougherty, Bittner, and Akay 2011; Braitenberg 2011; Davies 2019). Reality for a cognitive agent is an informational structure (Dodig-Crnkovic 2017; Dodig-Crnkovic and von Haugwitz 2017) and biology computes. Processes of change in informational structures establish computational dynamics. This model of reality for an agent includes both the information about the agent itself and about the world as it appears for the agent via interactions with the environment.

Proposed naturalist framework provides computational architecture for cognitive info communications which the author has been developing (Dodig-Crnkovic 2006; 2012; 2016; 2017; 2018; 2020). In the info-computational approach, evolutionary process unfolds in living organisms, and it happens in the sense of extended evolutionary synthesis (Laland et al. 2015; Ginsburg and Jablonka 2019; Jablonka and Lamb 2014; Gontier 2010) as a result of interactions (communication) between natural agents, be it cells, their groups or multicellular organisms.

The difference to the "Modern evolutionary synthesis", which supplemented Darwinism with added genetic determinism, in the computing nature approach, the emphasis is on *the role of the interaction/communication with the environment for the development and evolution*. Origins of life can be found in the first simplest pre-biotic chemical agents, leading to more complex forms such as viruses and furthermore first cells as bacteria, continuing up in complexity through the information self-organization. Genes are important, but not the solely responsible for the development of cells and their aggregates up to organisms and ecologies. As (Ginsburg and Jablonka 2019; Jablonka and Lamb 2014; Witzany and Baluška 2012a) describe, the interplay between the genetic code, through material embodiment, with the environment is crucial. Even (Rovelli 2018) argues for the central role of *evolution* as a mechanism of generating *mental* (intentionality, purpose, agency) from *physical*: "Meaning and Intentionality = Information + Evolution".

Reflected in the framework of info-computational nature, living organisms are cognitive agents, from single cells to humans, (Dennett 2017; Dodig-Crnkovic and von Haugwitz 2017). Cognitive artifacts can also be seen as natural physical systems with various degrees of cognitive capacities (Almér, Dodig-Crnkovic, and von Haugwitz 2015; Dodig-Crnkovic 2014b). Cognition is an open-ended process of self-organization where computation proceeds as signal processing at physical and chemical levels, while on the biological and cognitive levels it takes form of symbol manipulation and language-based communication, (Ehresmann 2012; 2014).

Simultaneous development of minds and bodies has been studied by (Schroeder 2013) as natural information processes. In a cognitive agent, variety and its dynamics is tackled through dual concept of selective and structural aspects of information. Biological evolution of species is seen as dynamical information processing. What for a cognitive agent appears as "the world" is *an interface*, a shared boundary across which the information is exchanged (Rössler 1998), with perception based on data/information obtained from the senses. Similar idea of computational boundary of a "self" is put forward by (Levin 2019) who describes mechanisms driving biological agents towards multicellularity and scale-free cognition.

## 3    NATURAL INFO-COMPUTATION PROCESSES

In our study of cognition as natural phenomenon we adopt Informational structural realism (Floridi 2003; 2008) as an approach in which reality is "informational structure for an epistemic agent interacting with the universe by the exchange of data as constraining affordances." The dynamics of that informational structure is conceptualized through the idea of Computing nature - where the physical dynamics of informational nature is seen as natural computation (Rozenberg, Bäck, and Kok 2012) (Dodig-Crnkovic 2010; 2015; Burgin and Dodig-Crnkovic 2011; Dodig-Crnkovic and Giovagnoli 2013) (Bull et al. 2013). This provides a unified naturalist setting for studying structures and processes in both animate and inanimate world (Dodig-Crnkovic 2007). The underlying fundamental property of information which makes it suitable as a basis of naturalist approach is: "there can be no information without physical implementation" (Landauer 1996).

Our approach is based on naturalized ontology/metaphysics. As (Ladyman et al. 2007) explain, naturalists construct science-based, structural-realist, computational ontology. Physicist Rovelli contributes to the naturalization program with his proposal to build the *foundation of physics on relative information* (Rovelli 2015). Here Shannon's relative information between two physical systems defines a *purely physical notion of information*, which can be used to "glue everything together" (Dodig-Crnkovic 2012b). That means to connect networks of networks of information processing nodes in nature. Interactions between physical systems are exchanges of information that establish physical correlations between them, through Shannon's relative information (correlation). By combining physical correlations with Darwinian evolution, Rovelli builds a ground for emergence of meaning.

Insights from the field of informational chemistry helps further in bridging the gap between physics and biology, with supramolecular chemistry that connects molecular recognition, molecular information processing and self-organization (Lehn 2015; 2017). Biology and cognitive sciences are already established as information-based and their processes have been modelled as natural computation (Miller 2018; Igamberdiev 2017; Torday and Miller 2020; Forrest and Mitchell 2016). Even the evolution of life has been modelled as a process of morphological info-computation (meta-morphogenesis) (Sloman 2013).

The whole sequence of information-based sciences – from physics to chemistry, biology (including evolution of species), and cognitive sciences (including social cognition) makes it possible to understand cognition in its context of natural information processing/natural computation.

## 3.1 Natural/ Physical/ Morphological Cognitive Computation

The concept of natural computation as presented by (Rozenberg, Bäck, and Kok 2012; Dodig-Crnkovic 2013b) addresses information processing (both discrete and continuous), as spontaneously appearing in nature (Crutchfield, Ditto, and Sinha 2010). Models of natural computation/natural information processing differ from the Turing model of computation, that is symbol manipulation. From the point of view of organization of computational processes, natural computation is different from von Neumann computation. Natural computation models of biological organisms with their multi-level processes of computing and distributed information communication are capable of capturing the dynamic behavior of natural cognitive agents, including neuronal networks (Buzsáki 2009).

In the framework of info-computational nature, the fundamental mechanism is *morphological computation*, i.e. a process of information self-organization such as described in (Haken 2006; Haken and Portugali 2017; Haken 2008). The author addressed this topic in (Dodig-Crnkovic 2013b; 2017; 2018). Morphological computing in living nature is a network of morphological informational processes for cognitive agents, with cognition as layered morphological computation.

Recently, in robotics, specific use of the term "morphological computation" has been adopted to denote decentralized embodied control of robots. In the context of robotics, appropriate body morphology is saving information processing (computational) resources as well as enabling learning through self-structuring (self-organization) of information in a cognitive agent (Pfeifer, Iida, and Gomez 2006; Pfeifer, Lungarella, and Iida 2007; Hauser, Füchslin, and Pfeifer 2014). This is macroscopic view of morphological computation that do not concern lower levels of organization such as cellular, molecular or quantum computation.

Natural computation appears on all levels of organization in nature, from physical, chemical, biological to cognitive. Of special interest for us are the levels with chemical and biological computation contributing to cognitive behavior such as presented by (Lones et al. 2013), showing biochemical basis of connectionism. On the level of cells and tissues there are numerous computational approaches such as (Tyrrell et al. 2016; Turner et al. 2013; Lyon et al. 2021). Proposals have been made for synthetic analog computation of living cells with artificial epigenetic networks (Daniel et al. 2013). Extensive literature exists on specific neuronal computation (Piccinini 2020; Piccinini and Shagrir 2014), as well as computational models of brain, (Laughlin and Sejnowski 2003; Buzsáki and Draguhn 2004; Averbeck, Latham, and Pouget 2006; Sardi et al. 2017; Hertz, Krogh, and Palmer 2018; Barron et al. 2020).

It is important to keep in mind the difference between new computational models of intrinsic information processes in nature (natural computing/morphological computing), and old computationalism based on computer metaphor of the Turing machine, performing symbol processing, that has been rightly criticized as inadequate model of human cognition (Miłkowski 2018; Scheutz 2002).

As already pointed out, in humans there are two basic cognitive systems, System 1 (reflexive, non-conscious, automatic, intuitive information processing, which is fast) and System 2 (reflective, conscious, reasoning and decision making, which is slow) (Kahneman 2011; Tjøstheim et al. 2020). Recognizing only symbolic information processing leaves the symbol grounding problem unsolved. Sub-symbolic data/signal processing provides mechanisms of symbol grounding in deep learning.

Hybrid symbolic-dynamical architectures (Larue, Poirier, and Nkambou 2012; Bekkum et al. 2021) have been proposed as well, capable of modelling a combination of the two as a reactive-deliberative behavior. According to (Ehresmann 2014), the fast reflexive System 1 can be understood in terms of Rovelli's physical correlations (Shannon's relative information), and it can accommodate for emotion as argued in (von Haugwitz, Dodig-Crnkovic, and Almér 2015), while the slow System 2, because of synonymity in the symbol system, introduces element of choice and indeterminism with higher computational demands. The latter has been addressed in (Mikkilineni 2012), also addressing the topic of parallel concurrent computation typical of biological systems, for which the Turing Machine model is not adequate.

## 4    NATURALIZING COMMUNICATION

### 4.1    Computation vs Communication

Computation as well as communication involve the transition, transformation and preservation of information (Dodig-Crnkovic 2008). The relationship between communication and computation has been described by (Bohan Broderick 2004) who argues that they are not conceptually distinguishable. The only difference is that computation concerns actions within a system, while communication is a process of interaction between a system and its environment. Biological systems are open information processing systems in communication with the environment, where the boundary between the system and the environment is dynamic and blurred.

As all other fundamental concepts which are objects of intense research (including information, computation, and cognition), the concept of communication has no generally accepted definition (Tjøstheim et al. 2020). We will use the concept in the sense of computation between systems (Bohan Broderick 2004), that is as exchange of information between the system and the environment.

One often thinks of communication as being defined by language with addition of other symbol systems such as images and sounds. But if we think of a human being with all its senses, then communication takes place on many different levels and through many channels that are interacting in the brain. For example, one does not think and feel clearly when disturbed by a constant noise, feeling strong anxiety, or is upset. The whole person participates in communication. Emotional signs may not always be as obvious as the conventional symbolic message, but can be orders of magnitude more important for the receiver, like bodily language, tone of voice, smile, eye gaze etc. (Damasio 1999; von Haugwitz, Dodig-Crnkovic, and Almér 2015; Dodig-Crnkovic 2017). To understand communication in a more multifaceted way than the exchange of symbols or signs, it is instructive to look at communication in other, simpler organisms. All living beings communicate. A cell that is the basic building block of life is a complex communication system. Without communication, life would not be possible (Maldonado 2016).

Naturalized understanding of communication is based on information as being defined by structures and computational processes in nature (Burgin and Dodig-Crnkovic 2020). In the same way as epistemology (knowledge) can be naturalized (Dodig-Crnkovic 2007; 2010) using natural information processing (computation), even communication and cognition with intelligence can be naturalized.

**4.2   Relation to the Research Field of Cognitive Infocommunications (CogInfoCom)**

Connecting engineered and natural cognitive systems is a strong trend worth further exploration. Cognitive info-communications (CogInfoCom) is a field that links info-communications with the cognitive sciences, with engineering applications. "The goal of CogInfoCom is to provide a systematic view of how cognitive processes can co-evolve with info-communications devices so that the capabilities of the human brain may not only be extended through these devices, irrespective of geographical distance, but may also interact with the capabilities of any artificially cognitive system." (Baranyi, Csapo, and Sallai 2015) Through this unification of cognitive capabilities of engineering artifacts and natural cognitive systems new qualities of information communication and cognition are achieved (Baranyi, Csapo, and Sallai 2015).

**5   TIME ASPECT OF COGNITIVE ARCHITECTURE, LEARNING AND MEMORY IN NATURALIZED COMMUNICATION**

Typically, cognitive architectures assume the mind/brain to be reactive, where information processing starts with a stimulus and ends in a response (Bechtel 2013). However, cells are inherently active, neurons are sustained oscillators, exhibiting electrochemical oscillations even in the absence of stimuli. Input data/information presents stimuli that *modulate existing endogenous oscillations*. (Bechtel 2013). In the book "Rhythms of the Brain" (Buzsáki 2009) describes the important role that spontaneous activity of neurons plays. Spontaneous firing of neurons is the very basis of human cognition when it comes to its time aspects. A self-organized timing of oscillations has co-evolved as the main organizational principle of neuronal activity. Global computation (on multiple spatial and temporal scales) is enabled by small-world-connectivity of neurons in the cerebral cortex. In a small-world setting, any two of nodes are connected through a short sequence of intermediary nodes. Cortical system is in a metastable state, synchronized through weak links between network oscillations in constant interactions. Oscillator frequency determines periods of receiving and transferring information.

Based on studies of oscillations, neural computations and learning, (Penagos, Varela, and Wilson 2017) propose that "*precisely coordinated representations across brain regions allow the inference and evaluation of causal relationships to train an internal generative model of the world.*" Training starts while awake, and processing continues during sleep when periodic nested oscillations induce hierarchical processing of information. Authors suggest that "general inference, prediction and insight" supporting an internal model for generalization and adaptive behavior is enabled through periodic states of sleep.

Related is the synaptic plasticity of the brain which changes its connections through the long-term potentiation (Hebbian and non-Hebbian), considered to be a basis for learning and memory. Oscillatory behavior is not only characteristics of the human brain. Similar oscillatory rhythms have been observed in the brains of mice. Being made of *oscillators, biological neural networks are able to filter inputs and to resonate with noise.* Unlike those observed oscillatory time behaviors in the biological brains, that appear as a result of their physical embodiment, artificial neural networks have no such temporal coupling and synchronizing mechanisms. It is an open question how essential this oscillatory behavior and metastability are for "fine tuning to the world" and if their function can be obtained in a different way.

On the global level of unified theories of cognition, time aspect (Anderson 2002) manifests itself in terms of Newell's *bands of cognition* (Newell 1994)—the biological "10 millisecond band", cognitive, rational, and social ("long-term") bands. How important is it to have all of them represented and how detailed? Here we talk about understanding of temporal aspects of cognition as organized hierarchically in a metastable state, constantly tuning to the environment. Coordination obtained through communication is central for connecting different levels, from molecules to thoughts, in the same coordination dynamics (Kelso, Dumas, and Tognoli 2013). Through the interplay with the environment this process results in *eigenstates* (Foerster 2003). Technological approaches to cognitive architecture of brain-like computer, based on frequency-fractal computing are proposed by (Ghosh et al. 2014) and (Singh et al. 2020).

## 6 DISCUSSION. OPEN QUESTIONS OF COGNITIVE ARCHITECTURES IN THE LIGHT OF NATURAL INFO-COMPUTATION

With the present development of cognitive and intelligent computing it is becoming important to improve computational approaches to cognitive architectures. Currently, there are several interesting open questions worth more exploration.

### 6.1 Biomimetic Design of Cognitive Architectures. How "Biologically Plausible" is Enough?

Proposals to learn from nature about cognition are not new, but they have recently gained a lot of prominence in the form of biomimetic design, such as (Joyee et al. 2020). Can our newly acquired insights into cognition on different levels of organization in nature be applied to design of cognitive architectures?

(Russin, O'Reilly, and Bengio 2020) suggest that deep learning, which has been inspired by information processing/ computation by neuronal networks in the brain, corresponds to Kahneman's

System 1 (fast, reflexive information processing). It needs an equivalent of "prefrontal cortex" that would play the role of System 2 (slow, reflective information processes). This is in agreement with (Marblestone, Wayne, and Kording 2016) who suggest increased integration of deep learning and neuroscience. Similar ideas are put forward by (Dodig-Crnkovic 2020) with the arguments that natural morphological computation should be used to study function of meta-learning (learning to learn) in humans (function of prefrontal cortex), other living organisms, and intelligent machines. Learning from nature and biomimetic design necessitate interdisciplinary approaches to neural computing (Esposito et al. 2018), emerging from research in variety of research fields.

## 6.2 Computational Efficiency and Natural Computing

Computational efficiency and performance are important features, often left outside when discussing computational models of cognition. However, with the increased ubiquity of computing, this aspect becomes essential. Natural cognitive computing can provide ideas for future developments towards more resource-efficient computational architectures (Usman et al. 2019) (Nature Editorial 2019).

Synthetic biology is one of the fields where efficient computation in the resource-limited environment of cells is essential for diagnostic, therapeutic and technological applications. For example, (Daniel et al. 2013) engineered synthetic analog gene circuits for advanced computational functions of multi-signal integration and processing in living cells, with only three transcription factors.

The question of computational efficiency has also been addressed by biomimetic neuromorphic computing which is mimicking the neural structure and functions of the human brain, together with probabilistic computing, with algorithmic approaches to the uncertainty, ambiguity, and contradiction in nature (Ackerman 2019). More learning from nature about computational efficiency is needed that will inform biomimetic designs of cognitive architectures.

## 6.3 Cognitive Behaviors and their Simulation, Emulation and Engineering

In the special report "Can We Copy the Brain?" (The Editors of IEEE Spectrum 2017), the founder of the Blue Brain Project, Henry Markram discusses complexities of the brain and necessity of learning about the details of its functioning on different levels of organization. He also discusses possibility to simulate the brain with molecular and cellular level simplified and encapsulated. Those are two open questions that run in parallel, providing an opportunity for two-way learning between computing and neuroscience (Rozenberg and Kari 2008). The questions are: first, how cognition works and develops in nature, and second, how we can simulate, emulate and engineer it.

Work of Michael Levin (https://ase.tufts.edu/biology/labs/levin/) suggests broad range of applications for nature-inspired cognitive architectures based on biological cognition connecting genetic networks, cytoskeleton, neural networks, tissue/organ, organism with the group (social) levels of information processing. Levin shows how biology has been computing through somatic memory (information storage) and biocomputation/decision making in pre-neural bioelectric networks, before the development of neurons and brains. Insights from biocognition can help the development of new AI platforms, applications in targeted drug delivery, regenerative medicine and cancer therapy, nano-technology, synthetic biology, artificial life, and much more.

# 7 CONCLUSION

This mini-review presents emerging advances in understanding of cognitive communication processes in nature, that can be used to inform future bio-inspired/biomimetic cognitive architectures in the range of applications, from nano-technology to medicine and robotics. Interpreting the nature in terms of computation (information processing), we can better understand processes of cognition as they function and evolve in living beings, from single cells to complex organisms, and their networks. Computation in nature/natural computation/physical computation/morphological computation stands for processes of self-structuring of information in a number of organizational levels: physical, chemical, biological, cognitive, and social with networks of communicating agents on every level of organization (Dodig-Crnkovic 2017). It is important to note the parallel development of our understanding of cognition as natural phenomenon and its technological implementations that inform each other in a recursive manner (Rozenberg and Kari 2008)(Bondgard and Levin 2021). At the same time as the knowledge of cognitive processes in living organisms increases, so do also our information processing/computational models.

We started this review by the description of the von Neumann computational architecture, based on the understanding of the brain from 1958. A decade later, visionary anticipating the convergence between biology and computing, von Neumann proposed bio-inspired theory of self-reproducing automata (von Neumann 1966). Development towards biomimetic architectural design promises new resource-effective cognitive architectures on different levels of complexity – from basal cognition useful in nanotechnology to complex cognition needed for social robotics. A lot of research is currently being done and the field is still in its infancy.

Starting with the overview of the existing cognitive architectures as presented in (Kotseruba and Tsotsos 2020) we are pointing to the fact that biologically inspired models of cognition such as new developments of connectionist (hybrid deep learning) and dynamical models, including non-neural cognitive systems deserve place among cognitive architectures.

# 8 Conflict of Interest

*The author declares that the research was conducted in the absence of any commercial or financial relationships that could be construed as a potential conflict of interest.*

# 9 Author Contributions

This is a work of a single author.

# 10 Funding

Research supported by the Swedish Research Council project MORCOM@COG Morphological Computing in Cognitive Systems Morcom (2016-2021).

# 11 Acknowledgments

The authors want to acknowledge the valuable comments and suggestions by three anonymous reviewers. Furthermore, the insightful suggestions by Zoran Konkoli and Gerard Jagers op Akkerhuis are gratefully acknowledged.

Natural Computational Architectures

Natural Computational Architectures